\documentclass[twocolumn,english,notitlepage,aps,10pt]{revtex4-1}
\usepackage[T1]{fontenc}
\usepackage[latin9]{inputenc}
\usepackage{amsmath}
\usepackage{amssymb}
\usepackage{graphicx}

\makeatletter
\usepackage{fancyhdr,lipsum}
\usepackage[colorlinks,linkcolor=blue,anchorcolor=blue,citecolor=blue, urlcolor=blue]{hyperref}
\usepackage{times}
\usepackage{tikz}

\makeatother

\usepackage{babel}
\begin{document}
\title{Isomeric excitation of $^{229}\mathrm{Th}$ via scanning tunneling microscope}
\author{Xue Zhang}
\address{Graduate School of China Academy of Engineering Physics, No. 10 Xibeiwang East Road, Haidian
District, Beijing, 100193, China}
\author{Tao Li}
\address{Graduate School of China Academy of Engineering Physics, No. 10 Xibeiwang East Road, Haidian
District, Beijing, 100193, China}
\author{Xu Wang}
\email{xwang@gscaep.ac.cn}

\address{Graduate School of China Academy of Engineering Physics, No. 10 Xibeiwang East Road, Haidian
District, Beijing, 100193, China}
\author{Hui Dong}
\email{hdong@gscaep.ac.cn}

\address{Graduate School of China Academy of Engineering Physics, No. 10 Xibeiwang East Road, Haidian
District, Beijing, 100193, China}
\begin{abstract}
The low energy of the isomeric state of the radionuclide thorium-229 ($^{229}\mathrm{Th}$)
makes it highly promising for applications in fundamental physics, precision metrology, and
quantum technologies. However, directly accessing the isomeric state from its ground state
remains a challenge. We propose here a tabletop approach utilizing the scanning tunneling microscope
(STM) technique to induce excitation of a single $^{229}\mathrm{Th}$ nucleus. With achievable
parameters, the isomeric excitation rate is advantageous over existing methods, allowing the
excitation and control of $^{229}\mathrm{Th}$ on the single-nucleus level. It offers the unique
potential of exciting and detecting subsequent $\gamma$ decay from a single nucleus, providing
a new direction for future experimental investigation of the $^{229}\mathrm{Th}$ isomeric
state. 
\end{abstract}
\maketitle
\textit{Introduction.---} $^{229}\mathrm{Th}$ recently garnered significant attention due
to its low-lying isomeric state \citep{Kroger1976,Reich1990}, which is only 8.3 eV above the
nuclear ground state \citep{Seiferle2019,Kraemer2023}. It is an appealing candidate for various
applications in constructing nuclear optical clocks \citep{Peik2003,Rellergert2010,Campbell2012,Beeks2021},
detecting temporal variations of fundamental constants \citep{Flambaum2006,Berengut2009,Kazakov2012},
measuring gravitational shifts \citep{Ludlow2015,Safronova2016}, etc. The isomeric state can
be obtained from nuclear decay reactions \citep{Barci2003,Thielking2018,Kraemer2023}. Nevertheless,
to allow control and to facilitate the applications, extensive research efforts have been made
to explore active nuclear-excitation approaches, using vacuum ultraviolet light sources \citep{Jeet2015,Yamaguchi2015,Stellmer2018},
high-energy synchrotron radiations \citep{Tkalya2000,Masuda2019}, laser pulses \citep{Tkalya1992,Borisyuk2018excitation,Porsev2010,Nickerson2020,Wang2021,Qi2023},
electrons \citep{Tkalya2020,Zhang2022}, muons \citep{Tkalya2021,Garguilo2022}, etc. Currently,
experimental demonstrations are only reported with high-energy synchrotron radiations \citep{Masuda2019}
and laser-generated plasmas \citep{Borisyuk2018excitation}. More manipulable experimental
approaches are still desirable.

\begin{figure}
\begin{centering}
\includegraphics{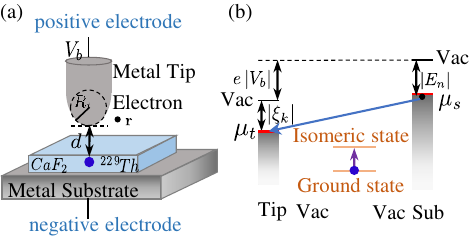}
\par\end{centering}
\caption{Schematic illustration of STM with a single $^{229}\mathrm{Th}$ atom located in $\mathrm{CaF_{2}}$.
The $^{229}\mathrm{Th}$ atom is doped in a $\textrm{Ca\ensuremath{\textrm{F}_{2}}}$ crystal
to suppress internal conversion. The STM tip apex is modeled as a sphere with radius $R_{t}$.
The position of the $^{229}\mathrm{Th}$ atom (blue dot), just below the center of the tip,
is set as the origin of the coordinate system, and $d$ is the distance between the tip and
substrate. $\mathbf{r}$ stands for the position of the tunneling electron (black dot), and
$V_{b}$ is the bias voltage applied to the tip and substrate. (b) The energy level diagram
at negative bias voltage. The black lines denote the vacuum level for two electrodes, and the
red lines represent the initial and final electronic states. $\mu_{t}\equiv\mu_{0}+eV_{b}$
and $\mu_{s}\equiv\mu_{0}$ are the Fermi energies of the tip and the substrate at the bias
voltage $V_{b}$, where $\mu_{0}$ is the Fermi energy of the tip and the substrate at zero
bias.}

\label{fig:system}
\end{figure}

In this Letter, we propose a completely new experimental setup using the tabletop scanning
tunneling microscopy (STM) to excite the $^{229}\mathrm{Th}$ atomic nucleus to its isomeric
state $^{229m}\mathrm{Th}$. STM is a powerful imaging technique used in nanotechnology and
surface science \citep{Binnig1982,Tersoff1983}. It relies on the principle of quantum tunneling,
where a sharp metal tip scans the surface of a sample at atomic scales, detecting the flow
of electrons between the tip and the surface. By mapping the electron tunneling current, STM
produces high-resolution images that reveal the topography and electronic properties of materials
at the atomic level. With the high spatial control of the tip, a single $^{229}\mathrm{Th}$
atom can be located and controlled, as illustrated in Fig. \ref{fig:system}. A metal tip with
radius of curvature $R_{t}$ is positioned above the substrate plane with distance $d$. Both
the tip and substrate are typically made of a noble metal, e.g. silver (Ag) \citep{Hla2005,Oka2014}.
The $^{229}\mathrm{Th}$ atoms are assumed to be doped in a wide-bandgap crystal, such as $\textrm{Ca\ensuremath{\textrm{F}_{2}}}$,
to suppress the internal conversion process \citep{Dessovic2014}. After applying a bias voltage
$V_{b}$, electrons will tunnel through the vacuum between the tip and the substrate and excite
the $^{229}\mathrm{Th}$ nucleus from the ground state to the isomeric state.

The current approach offers advantages on precise manipulation, isomeric-excitation efficiency,
and photon-detection efficiency. Firstly, the STM allows the precise focus of electronic current
on the level nA to an area of nm scale \citep{Eigler1990}, resulting in an electron flux of
about $10^{10}$ $\mathrm{nm^{-2}s^{-1}}$. In contrast, beam electron sources have current
intensities on mA level and cm-scale area \citep{Borisyuk2016}, yielding an electron flux
of $10^{2}$ $\mathrm{nm^{-2}s^{-1}}$. This key difference allows (a) excitation and control
of a located single $^{229}\mathrm{Th}$ nucleus, and (b) much higher single-nucleus excitation
rate. With currently available STM parameters (e.g. tip radius 0.5 nm, tip-substrate distance
0.5 nm, and bias voltage $-11$ V), the isomeric excitation rate can reach $10^{-5}$$\mathrm{s}^{-1}$.
Further technological refinement may increase it to the level of $10^{-2}$$\mathrm{s}^{-1}$,
opening enormous potential in single-nucleus excitation and control. Secondly, the current
tabletop setup enables the application of the experimental technique of the highly efficient
luminescence detection with the detection solid angle about 3 sr \citep{Zhang2015,Berndt1991},
which is much greater than the solid angle of about 0.1 sr in the synchrotron radiation excitation
experiment \citep{Masuda2019}. Efficient photon collection allows the detection of the weak
photon signals from the radiative decay of $^{229m}\mathrm{Th}$. These advantages make the
current method highly promising for achieving excitation, control, and detection of $^{229}\mathrm{Th}$
especially on the single-nucleus level, which is not achievable with other methods.

\textit{Theory of Isomeric Excitation using STM.---} In this part we develop a quantum theory
of isomeric excitation particularly for the STM setup. The total Hamiltonian of the system
is $H=H_{el}+H_{n}+H_{int}$, where $H_{el}(H_{n})$ represents the Hamiltonian of the tunneling
electron (the $^{229}\mathrm{Th}$ nucleus) and $H_{int}$ is the interaction between them.
The Hamiltonian of the tunneling electron is $H_{el}=-\nabla^{2}/2m_{e}+V(\mathbf{r})$, where
$V(\mathbf{r})$ is the potential felt by the tunneling electron at the position $\mathbf{r}$
\citep{Dong2020,Dong2021}. The wave functions are found for different regions as \citep{Bardeen1961,Gottlieb2006,Dong2020,Dong2021}
\begin{align}
H_{el,t}\left|\phi_{k}\right\rangle  & \simeq\tilde{\xi}_{k}\left|\phi_{k}\right\rangle ,\nonumber \\
H_{el,s}\left|\varphi_{n}\right\rangle  & \simeq\tilde{E}_{n}\left|\varphi_{n}\right\rangle ,
\end{align}
where $H_{el,t}$($H_{el,s}$) is the Hamiltonian of the free tip (substrate) obtained by neglecting
the potential in the substrate (tip) region. $\left|\phi_{k}\right\rangle $($\left|\varphi_{n}\right\rangle $)
is the eigenstate of free tip (substrate) with $\tilde{\xi}_{k}\equiv\xi_{k}+eV_{b}$ ($\tilde{E}_{n}\equiv E_{n}$),
where $\xi_{k}$($E_{n}$) is the eigenenergy with zero bias voltage. Here we neglect the change
of the wave function of the tip induced by the applied voltage \citep{Chen1990}. The detailed
forms of these wave functions are presented in the Supplementary Materials (SM).

The Hamiltonian of the $^{229}\mathrm{Th}$ atomic nucleus is simplified as a two-level system
$H_{n}=\epsilon_{g}\left|g\right\rangle \left\langle g\right|+\epsilon_{e}\left|e\right\rangle \left\langle e\right|$,
where $\left|g\right\rangle $($\left|e\right\rangle $) is its ground (isomeric excited) state
with energy $\epsilon_{g}$($\epsilon_{e}$). We assume that the energy gap is $\Delta_{eg}\equiv\epsilon_{e}-\epsilon_{g}=8.338\ \mathrm{eV}$
\citep{Kraemer2023}. The interaction between the electron and nucleus is given by \citep{Eisenberg1976}
\begin{equation}
H_{int}=-\frac{1}{c}\int\mathbf{J}(\mathbf{R})\cdot\mathbf{A}(\mathbf{R})d\mathbf{R},
\end{equation}
where $\mathbf{J}(\mathbf{R})$ is the nucleus current operator at position $\mathbf{R}$.
$\mathbf{A}(\mathbf{R})$ is the electromagnetic vector potential generated by the tunneling
electron with current $\mathbf{j}(\mathbf{r})$, 
\begin{equation}
\mathbf{A}(\mathbf{\mathbf{R}})=\frac{1}{c}\int\frac{e^{ik|\mathbf{r}-\mathbf{R}|}}{|\mathbf{r}-\mathbf{R}|}\cdot\mathbf{j}(\mathbf{r})d\mathbf{r},\label{eq:vector potential}
\end{equation}
where $k\hbar c=(\tilde{E}_{n}-\tilde{\xi}_{k})$ corresponds to the energy loss of the electron.
The electron current is obtained as $\mathbf{j}_{fi}(\mathbf{r})=-ei\hbar(\psi_{f}\nabla\psi_{i}^{*}-\psi_{i}^{*}\nabla\psi_{f})/2m_{e}$
\citep{Bardeen1961,Tersoff1983}. And $\psi_{i}$ ($\psi_{f}$) denotes the wave function of
the initial (final) state of the electron. For negative bias $V_{b}<0$, the electron flies
from the substrate to the tip, i.e. $\psi_{i}=\varphi_{n}(\mathbf{r})$ and $\psi_{f}=\phi_{k}(\mathbf{r})$.

From Fermi's golden rule, the transition probability per unit time from an initial state $\left|i\right\rangle $
to a final state $\left|f\right\rangle $ is written as 
\begin{equation}
P_{fi}=\frac{2\pi}{\hbar}\left|\left\langle f\right|H_{int}\left|i\right\rangle \right|^{2}\delta(\mathcal{E}_{i}-\mathcal{E}_{f}),\label{eq:probability}
\end{equation}
where for the current system $\left|i\right\rangle \equiv\left|J_{i}M_{i}\right\rangle \otimes\left|\varphi_{n}\right\rangle $
and $\left|f\right\rangle \equiv\left|J_{f}M_{f}\right\rangle \otimes\left|\phi_{k}\right\rangle $
are the product states of the nucleus and the electron. Here $J_{i}$ ($J_{f}$) and $M_{i}$
($M_{f}$) are the angular momentum and magnetic quantum numbers of the nuclear ground (isomeric)
state, respectively. The initial and final energies are $\mathcal{E}_{i}=\epsilon_{g}+\tilde{E}_{n}$
and $\mathcal{E}_{f}=\epsilon_{e}+\tilde{\xi}_{k}$.

Here we use multipole expansion \citep{Eisenberg1976}
\begin{equation}
\frac{e^{ik|\mathbf{r}-\mathbf{R}|}}{|\mathbf{r}-\mathbf{R}|}=4\pi ik\sum_{\mathcal{T},l,m}\mathcal{A}_{lm}^{\mathcal{T}}(kR)\mathcal{B}_{lm}^{\mathcal{T}}(kr).
\end{equation}
The transition type $\mathcal{T}$ can be either $E$ (electric) or $M$ (magnetic). And $\mathcal{A}_{lm}^{\mathcal{T}}(kR)$
is the multipole vector potential, $\mathcal{A}_{lm}^{M}(kR)=1/\sqrt{l(l+1)}\mathbf{L}j_{l}(kR)Y_{lm}(\hat{\mathbf{R}})$
and $\mathcal{A}_{lm}^{E}(kR)=-i/(k\sqrt{l(l+1)})\nabla\times\mathbf{L}j_{l}(kR)Y_{lm}(\hat{\mathbf{R}})$.
Here $Y_{lm}(\hat{\mathbf{R}})$ are spherical harmonics. The potential $\mathcal{B}_{lm}^{\mathcal{T}}(kr)$
can be obtained from $\mathcal{A}_{lm}^{\mathcal{T}}(kR)$ by replacing the Bessel function
$j_{l}(kR)$ with the Hankel function of the first kind $h_{l}^{(1)}(kr)$. Then the transition
matrix element $\left\langle f\right|H_{int}\left|i\right\rangle $ turns into \citep{Eisenberg1976,Tkalya2020,Zhang2022}
\begin{align}
\left\langle f\right|H_{int}\left|i\right\rangle  & =\sum_{\mathcal{T}}\left\langle f\right|H_{int}^{\mathcal{T}}\left|i\right\rangle \nonumber \\
 & =-\frac{4\pi ik}{c^{2}}\sum_{\mathcal{T},l,m}\int\mathbf{J}_{fi}(\mathbf{R})\cdot\mathcal{A}_{lm}^{\mathcal{T}}(kR)d\mathbf{R}\nonumber \\
 & \qquad\qquad\quad\quad\times\int\mathbf{j}_{fi}(\mathbf{r})\cdot\mathcal{B}_{lm}^{\mathcal{T}}(kr)d\mathbf{r}.\label{eq:Hintif}
\end{align}

The first integral in the above equation associated to the nuclear transition current $\mathbf{J}_{fi}(\mathbf{R})$
is derived with the following form \citep{Tkalya2020} 
\begin{align}
\int\mathbf{J}_{fi}(\mathbf{R})\cdot\mathcal{A}_{lm}^{\mathcal{T}}(kR)d\mathbf{R} & =\frac{ik^{l}c}{(2l+1)!!}\sqrt{\frac{l+1}{l}}\nonumber \\
 & \quad\times|\left\langle J_{f}M_{f}\right|\mathcal{M}_{lm}^{\mathcal{T}}\left|J_{i}M_{i}\right\rangle |,
\end{align}
where the nuclear transition matrix element is related to the reduced probability $B(\mathcal{T}l;J_{i}\rightarrow J_{f})$
of nuclear transition \citep{Tkalya2020}
\begin{equation}
B(\mathcal{T}l;J_{i}\rightarrow J_{f})=\frac{1}{2J_{i}+1}\sum_{M_{i},M_{f}}\left|\left\langle J_{f}M_{f}\right|\mathcal{M}_{lm}^{\mathcal{T}}\left|J_{i}M_{i}\right\rangle \right|^{2}.
\end{equation}
The second integral of Eq. (\ref{eq:Hintif}), $\Xi_{fi}^{\mathcal{T}l}\equiv\int\mathbf{j}_{fi}(\mathbf{r})\cdot\mathcal{B}_{lm}^{\mathcal{T}}(kr)d\mathbf{r}$,
is associated to the electronic transition and can be deduced into the following forms 
\begin{align}
\Xi_{fi}^{El} & \approx-\frac{eicl}{\sqrt{l(l+1)}}\int\phi_{k}(\mathbf{r})\varphi_{n}(\mathbf{r})h_{l}^{(1)}(kr)Y_{lm}(\hat{\mathbf{r}})d\mathbf{r},\\
\Xi_{fi}^{Ml} & =0,
\end{align}
where $\Xi_{fi}^{El}$ is given to the leading order according to the condition $kr\ll1$.
Detailed derivations and discussions on $\Xi_{fi}^{\mathcal{T}l}$ are presented in the SM.

The overall transition probability per unit time is obtained explicitly as 
\begin{align}
P & =\frac{2\pi}{\hbar}\left(\frac{4\pi k}{c}\right)^{2}\frac{k^{2l}}{\left[(2l+1)!!\right]^{2}}\frac{l+1}{l}\frac{B(El;J_{i}\rightarrow J_{f})\delta(\mathcal{E}_{i}-\mathcal{E}_{f})}{\int dE_{n}\rho_{s}(E_{n})}\nonumber \\
 & \quad\times\sum_{n,k}F_{\mu_{0},T}(E_{n})\left(1-F_{\mu_{0},T}(\xi_{k})\right)|\Xi_{fi}^{El}|^{2},
\end{align}
where $\rho_{s}(E)\,(\,\rho_{t}(E)\,)$ is the density of state at the substrate (tip). $F_{\mu_{0},T}(E)$
is the Fermi--Dirac distribution of electrons in tip or substrate state with chemical potential
$\mu_{0}$ and temperature $T$. In an STM experiment, the temperature of the ultrahigh-vacuum
chamber is low enough, typically lower than 10 K \citep{Binnig1987,Hansma1988}, that the Fermi--Dirac
distribution function is approximately a Heaviside function, i.e. $F_{\mu_{0},T}(E)=1$ for
$E<\mu_{0}$ and $F_{\mu_{0},T}(E)=0$ for $E>\mu_{0}$. The transition probability per unit
time is simplified as 
\begin{align}
P & =\frac{2\pi}{\hbar}\left(\frac{4\pi k}{c}\right)^{2}\frac{k^{2l}}{\left[(2l+1)!!\right]^{2}}\frac{l+1}{l}\frac{B(El;J_{i}\rightarrow J_{f})}{\int dE_{n}\rho_{s}(E_{n})}\nonumber \\
 & \qquad\times\int_{\mu_{0}+\Delta_{eg}+eV_{b}}^{\mu_{0}}dE_{n}\rho_{s}(E_{n})\rho_{t}(\xi_{k})\left|\Xi_{fi}^{El}\right|^{2},\label{eq:probability_result}
\end{align}
where $\xi_{k}=E_{n}-\Delta_{eg}-eV_{b}$. Without loss of generality, we consider the material
of the tip and the substrate to be Ag, whose density of states is obtained from Ref. \citep{Lin2008}
(see the SM for details). In the calculation, we use the reduced nuclear transition probability
$B(E2;J_{f}\rightarrow J_{i})=27.04$ W.u. \citep{Minkov2017}.

\begin{figure}
\begin{centering}
\includegraphics{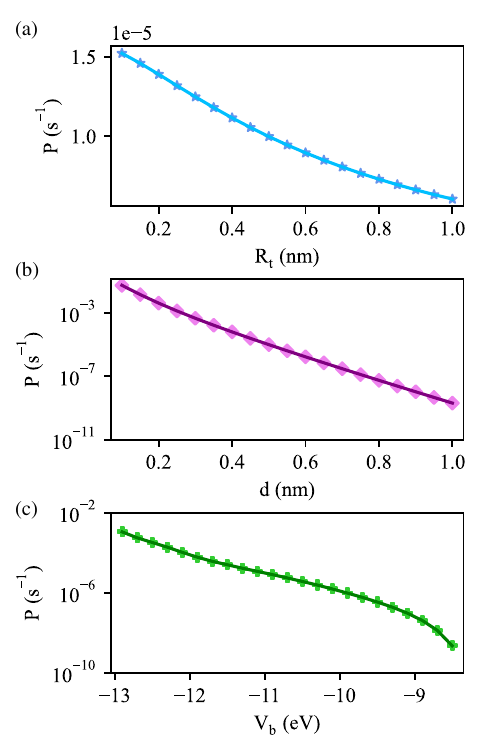}
\par\end{centering}
\caption{Transition probability per unit time $P$ as a function of tip radius $R_{t}$, distance from
the tip to the substrate $d$, and applied bias voltage $V_{b}$. (a) Dependency of $P$ on
$R_{t}$, with $d$ fixed at 0.5 nm and $V_{b}$ set to $-11$ V. (b) Dependency of $P$ on
$d$, with $R_{t}$ fixed at 0.5 nm and $V_{b}$ set to -11 V. (c) Dependency of $P$ on $V_{b}$,
with $R_{t}$ fixed at 0.5 nm and $d$ fixed at 0.5 nm. Symbols are numerical results, and
curves are added to guide the eye. \label{fig:result}}
\end{figure}

\textit{Numerical Results. ---} Fig. \ref{fig:result}(a) shows the transition probability
$P$ per unit time as a function of the tip radius $R_{t}$, for fixed tip position $d=0.5$
nm and applied bias voltage $V_{b}=-11$ V. The curve shows an exponential decay with the increase
of the tip radius. In Eq. (\ref{eq:probability_result}), the tip radius $R_{t}$ affects the
transition probability via the interaction strength $\Xi_{fi}^{El}$. A larger needle tip radius
leads to a weaker electric field strength at the tip, yielding a lower isomeric excitation
probability. And a smaller tip radius leads to a higher isomeric transition probability. For
small bias voltages, a non-monotonic dependency may appear though, and the detailed discussions
are presented in the SM.

Fig. \ref{fig:result}(b) presents the dependency of the transition probability $P$ on the
tip-substrate distance $d$, with $R_{t}$ fixed at 0.5 nm and $V_{b}$ set to $-11$ V. The
data shows that the transition probability $P$ decreases exponentially with the increase of
$d$. This is attributed to the decreasing overlap of wave functions between the tip and the
substrate as the distance increases. A larger wave-function overlap facilitates electron tunneling
between the tip and substrate, leading to an enhanced isomeric excitation probability.

Fig. \ref{fig:result}(c) shows an approximately exponential dependency of the transition probability
$P$ on the applied bias voltage $V_{b}$, in which both $d$ and $R_{t}$ are fixed at 0.5
nm. In Eq. (\ref{eq:probability_result}), the bias voltage $V_{b}$ mainly determines the
range of the energy integration. A higher $V_{b}$ expands the integration range, thus allowing
a broader range of energy levels to contribute to the isomeric excitation.

\textit{Discussions. ---} For a typical STM setup with tip radius $R_{t}=0.5$ nm, tip-substrate
distance $d=0.5$ nm, and bias voltage $V_{b}=-11$ V, the isomeric excitation rate is calculated
to be on the order of $10^{-5}$ s$^{-1}$. Here the voltage is chosen to be within the band
gap of the $\mathrm{CaF_{2}}$ crystal (11.6 $\sim$ 12.1 eV) to avoid crystal damage \citep{Dessovic2014}.
The tip-base distance $d=0.5$ nm is a typical parameter used in the STM experiments \citep{Stroscio1991,Kroeger2009}
to avoid the high electron currents.

We may compare the above rate to those of existing methods. (1) The indirect optical excitation
method using 29-keV synchrotron radiations yields an isomeric excitation rate on the order
of $10^{-11}$ s$^{-1}$ per nucleus \citep{Masuda2019}. (2) Isomeric excitation by inelastic
electron scattering is most efficient for electrons around 10 eV, and the corresponding cross
section is on the order of 1 mb, or $10^{-27}$ cm$^{2}$ \citep{Tkalya2020,Zhang2022}. Assuming
an electron beam with current 1 mA and beam area 1 cm$^{2}$, the electron flux is $6.25\times10^{15}$
cm$^{-2}$s$^{-1}$, and the isomeric excitation rate is about $6.25\times10^{-12}$ s$^{-1}$
per nucleus. The STM method is therefore advantageous in the excitation rate on the single-nucleus
level. This is mainly due to the capability of the STM to focus the tunneling electron current
on a nm-scale area.

The excitation rate has the potential to be further enhanced. Firstly, wider bandgap crystals
allow higher applied bias voltages hence higher excitation rates. Secondly, trying carefully
smaller tip-substrate distances (without burning out the crystal) allow higher excitation rates.
These attempts are obviously challenging, but they may enhance the excitation rate to an unprecedented
high level, for example, $10^{-2}$ s$^{-1}$ per nucleus. If so, together with the high photon
collection efficiency of STM, active excitation and subsequent detection of nuclear radiative
decay on the single-nucleus level may be realized. The quantum optical feature of nuclear $\gamma$
radiation could be carefully investigated.

\textit{Conclusion.---} In summary, we have proposed a new approach to use tunneling electrons
in STM to excite the thorium-229 nucleus from the ground state to the low-lying isomeric state.
The tunneling electrons, under an applied bias voltage, pass through the vacuum between the
tip and the substrate and excite the $^{229}$Th nucleus. A comprehensive theoretical framework
is developed to calculate the isomeric excitation rate, and to investigate the dependency of
the excitation rate on key STM parameters, including the tip radius, the tip-substrate distance,
and the applied bias voltage. The calculated single-nucleus excitation rate shows advantageous
over existing methods. More importantly, our method allows nuclear excitation and control on
the single-nucleus level, which is unique among all existing methods and proposals. The possibility
of exciting, controlling, and detecting nuclear radiative decay on the single-nucleus level
points to a completely new territory of studying nuclear physics as well as quantum optics.

This work is supported by the National Natural Science Foundation of China (NSFC) (Grants No.
12088101, No.U2230203, No. U2330401).

\bibliographystyle{apsrev4-1}
\bibliography{ref_nuclear,ref_STML}

\end{document}